\documentstyle[psfig,aps]{revtex}
\newcommand{\capt}[2]{
\begin{flushleft}
FIG. #1. #2
\end{flushleft}}

\newcommand{\An}{\AA$\;$}
\def\Deg{\hbox{$^\circ$}}
\begin{document}
\oddsidemargin -1cm
\topmargin -1.5cm
\textheight 25cm
\bibliographystyle{vm}

\title{First principles study of adsorbed Cu$_n$ (n=1-4) microclusters\\
on MgO(100): structural and electronic properties.}

\author{V. Musolino$^{1,2}$ \and A. Selloni$^{3,4}$ \and R. Car$^{1,2}$}

\address{$^1$Institut Romand de Recherche Num\'erique en Physique des
Mat\'eriaux (IRRMA),\\ PPH-Ecublens, CH-1015 Lausanne, Switzerland} 
\address{$^2$D\'epartement de Physique de la Mati\`ere Condens\'ee,
Universit\'e de Gen\`eve,\\ 24 quai Ernest-Ansermet, CH-1211 Gen\`eve
4, Switzerland} 
\address{$^3$D\'epartement de Chimie Physique, Universit\'e de
Gen\`eve,\\ 30 quai Ernest-Ansermet, CH-1211 Gen\`eve 4,
Switzerland}
\address{$^4$II Facoltà di Scienze Matematiche Fisiche e Naturali,
Università di Milano,\\ Sede di Como, I-22100 Como, Italy\vspace*{.5cm}}

\date{\today}

\maketitle

\begin{center}
{\bf Abstract}
\end{center}

\abstract{We present a density functional study  
of the structural and electronic properties 
of small Cu$_n$ ($n=1,4$) aggregates on defect-free  MgO(100).
The calculations employ a slab geometry with periodic
boundary conditions, supercells with up to 76 atoms, 
and include full relaxation of the surface layer and of all 
adsorbed atoms.
The preferred adsorption site for a single Cu adatom is on top of
an oxygen atom. The adsorption energy and Cu-O distance are
$E_{S-A} = 0.99$~eV and $d_{S-A} = 2.04$~\An using 
the Perdew-Wang gradient corrected exchange correlation functional.
The saddle point for surface diffusion is at the "hollow" site,
with a diffusion barrier of around 0.45 eV. 
For the adsorbed copper dimer, two geometries, one parallel and 
one perpendicular to the surface, are very close in energy.
For the adsorbed Cu$_{3}\;$, a linear configuration
is preferred to the triangular geometry.
As for the tetramer, the most stable
adsorbed geometry for  Cu$_{4}$ is  a rhombus. 
The adsorption energy per Cu atom decreases with increasing the
size of the cluster, while the Cu-Cu cohesive energy increases,
rapidly becoming more important than the adsorption energy.}

\section{Introduction}
\label{intro}

The interaction of metallic clusters with supporting 
metal-oxide surfaces is a subject of 
great current interest, because of  the numerous
technological applications of these systems, e.g. in the field of thin
film growth and catalysis\cite{ref:rev,ref:f}.
Important objectives of these studies are to understand
how the atomic and electronic structure of both subsystems
are modified through their interaction, as well as the properties of
the resulting interface.

In this paper we focus on the adsorption of small Cu$_n$($n=1,2,3,4$)
clusters on the non-polar (100) surface of MgO.
This surface has been widely investigated 
both experimentally and theoretically\cite{ref:rev}.
At present, all studies agree that the
atomic structure of the undefected surface is very
close to an ideally truncated bulk, even though,
from the electronic structure point of view, it is not
completely clear whether the reduced coordination at the surface gives
rise to a reduced ionicity and band gap with respect to the bulk.

Interest in metal clusters has grown remarkably
in the past few years. A major question has been to understand
the dependance of atomic and electronic structures on cluster size and
their evolution from the small cluster regime to the bulk. In
particular, copper clusters have been extensively studied, both
experimentally and theoretically. As Cu is characterized by a closed
d-shell and a single valence electron, an important issue has been to 
investigate similarities and/or differences between Cu$_{n}$ and simple
alkali-metal clusters. 
Theoretical work has been done at different levels
\cite{ref:mpc2,ref:ckrs,ref:j,ref:lb,ref:cj} 
and the properties of small Cu$_n$ ($n=1,5$) clusters are now quite
well understood. 

Studies of Cu clusters and/or overlayers on MgO(100) are not numerous.
While most studies conclude that the preferred Cu binding 
site is on top of a surface oxygen and that the Cu-surface
charge transfer is very small,
controversial results have been obtained on the fundamental issue of
the Cu-surface bond strength. 
Experimentally, measurements of the initial sticking probability
$S_0$ by medium energy ion scattering (MEIS)\cite{ref:zlg} have 
yielded $S_0 \sim 0.5$ at room temperature,
which has been interpreted in terms of weak
adsorption of Cu adatoms on the MgO surface. 
More recently, however, a substantially larger value of $S_0$, $\sim 0.82$,
has been obtained by thermal desorption(TD) techniques\cite{ref:wog}.
Theoretically, early Hartree-Fock calculations\cite{ref:bk} 
found that Cu binds strongly at Mg vacancy sites, while the binding 
with an undefected surface at the oxygen site is very weak (0.2 eV). 
By contrast, more recent Local Density Functional (LDF) calculations 
found a rather strong reactivity of the undefected surface, with 
an adsorption energy of 1.4 eV for an isolated Cu
adatom\cite{ref:llp} at the oxygen site. Similar LDF computations for
the Cu/MgO(100) interface\cite{ref:bmy} yielded a work of adhesion of
1.0 eV. However, gradient-corrected density functional cluster
calculations\cite{ref:pr} found that the binding of a single Cu atom
at the oxygen site is about 0.3 eV. 
It appears that the uncertainty in the value of the adsorption energy
is so large that even the character of the Cu-surface bonding is unclear.
In fact, while adsorption energies of the order of 1 eV or larger
suggest a (weak) covalent bond, a polarization mechanism should be
primarily involved in the case of binding energies of the order of
0.2-0.3~eV. 

In this paper we address this issue by means of 
first principles density functional calculations
within the framework of the Car-Parrinello approach \cite{ref:cp}.
Our calculations employ a slab geometry with periodic boundary conditions,
and, in order to avoid spurious interactions between 
periodic images of the adatom(s),
large supercells (up to 76 atoms) are considered.
Moreover, full relaxation of the surface layer and of the adsorbed
atoms is allowed for (without symmetry constraints). 
Our results show that theoretical predictions of the Cu-MgO binding energy 
based on density functional calculations are quite sensitive
to the choice of the exchange-correlation functional. 
On the basis of our most reliable calculations, which use
a gradient-corrected functional known as the Generalized
Gradient Approximation\cite{ref:gga}, the binding energy
for a Cu adatom on top of an oxygen of the undefected surface
is close to 1.0 eV.
Moreover analysis of the electronic charge distribution
confirms the occurrence of a weak covalent bond
between the Cu$3d/4s$
states and the O$2p$ orbitals.
The barrier for the Cu adatom to jump between neighboring binding
sites is $\sim 0.45$ eV, suggesting a substantial
mobility of the adatom at room temperature. 

Well defined trends show up in the dependence of energetic, structural
and electronic properties on the size of the adsorbed cluster.
In particular we find that the adsorption energy per atom decreases
with increasing the size of the cluster, while the Cu-Cu cohesive
energy increases, and large adsorbed clusters are energetically more
stable than smaller ones. 
Thus Cu$_n$ clusters do not dissociate, but rather tend to maintain
their identity when adsorbed on the surface. 
We also infer that Cu atoms which are deposited on the surface
at room (or higher) temperature
by thin-film growth techniques will tend to aggregate
and readily form larger clusters.
Moreover, calculations for Cu$_{5}$ and larger adsorbed clusters (to
be reported in detail elsewhere\cite{ref:m_else}) indicate a clear
preference for 3D over 2D geometries.
In this context, it may be interesting to remark
that experiments\cite{ref:hm2,ref:am} indicate that the growth
mechanism  for Cu on MgO(100) is Stranski-Krastanov, 
i.e. three-dimensional islands are present on top of a full monolayer. 

The layout of this paper is the following. 
In section \ref{comp} we give the details of our calculations. 
In section \ref{struc_ener}, after presenting some convergence tests, 
we discuss atomic structures and energetics.
Section \ref{elec} deals with electronic properties.
Both these sections are divided into three main parts. 
We first discuss the clean MgO(100) surface 
(sections \ref{struc_ener.clean}, and \ref{elec.clean}),
we then consider the free clusters
(sections \ref{struc_ener.free} and \ref{elec.free}), and
finally we deal with the adsorption of copper clusters on the surface in
sections \ref{struc_ener.supported} and \ref{elec.supported}. 
Both sections \ref{struc_ener} and \ref{elec} 
end with comments on observed trends
in the structural/energetical (\ref{struc_ener.supported.trends}) 
and electronic (\ref{elec.supported.trends})  properties.
A brief summary and conclusions are given in section \ref{conc}.

\section{Computational details}
\label{comp}

Calculations were carried out within 
the Car-Parrinello\cite{ref:cp} approach
using both the Local Density (LDA)\cite{ref:lpclv,ref:plclv}  
and the Generalized Gradient (GGA)\cite{ref:gga} approximations.
For a few calculations, we used also the so-called 
Becke exchange-only functional\cite{ref:b,ref:b1},
where gradient corrections are applied only to the exchange.
For the LDA exchange-correlation energy, we employed 
the Perdew-Zunger parametrization\cite{ref:pz}
of Ceperley and Alder's\cite{ref:ca} electron-gas results. 
For the GGA exchange-correlation energy, we used the 
functional given in Ref.\onlinecite{ref:gga}. 
Unless mentioned, the GGA was applied perturbatively, 
i.e. the total energy was calculated with the LDA charge density.
This non-self-consistent approach to GGA calculations has been shown
to yield energies which are usually in excellent 
agreement with fully-self-consistent GGA calculations 
(see, e.g., Ref.\onlinecite{ref:gga}). 

Valence-core interactions were described via Vanderbilt\cite{ref:v} 
pseudopotentials for O and Cu and
Bachelet-Hamann-Schl\"{u}ter\cite{ref:bhs} in the
Kleinmann-Bylander\cite{ref:kb} form for Mg. 
Pseudopotential cutoff radii were 1.5 and 2.0 a.u. for O and Cu respectively.
For Cu, valence electrons included the 3d and 4s shells.
Pseudopotentials were always consistent with the approximation used in
the density functional calculations, i.e. GGA pseudopotentials for
self-consistent GGA and LDA pseudopotentials otherwise (see
e.g. Ref. \onlinecite{ref:cpbc}). The smooth part of the electronic
wave-functions 
was expanded in plane-waves with a cut-off of 16 Ry.
Tests using a higher cutoff of 20 Ry were performed (see below). 
A cutoff of 150 Ry was used for the augmented electron
density\cite{ref:lpclv,ref:v}. 
As the supercells that we used were quite large,
the k sampling was restricted to the $\Gamma$ point.

The simulated systems were enclosed in cubic or orthorhombic cells of sizes 
ranging from 6 to $\sim 20$ \An with periodic boundary conditions (PBC). 
In all our calculations, we first performed an electronic minimization 
with steepest-descent and/or damped dynamics algorithms to bring
the electrons in the ground state corresponding to a given initial 
atomic configuration. We then relaxed the ions 
with a coupled electronic and ionic damped dynamics. 
We used the preconditioning scheme of Tassone {\em et al.}\cite{ref:tmc} 
to increase the simulation timestep. 
Timesteps were in the $1.0-1.4 \cdot 10^{-4}$ ps range with
a fictitious electronic mass $\mu= 1000$ a.u.. 

As a first test of the accuracy of our computational scheme, 
we calculated the equilibrium distance ($d$) and 
the vibrational frequency ($\nu$) of a few molecules
relevant to our study, namely Cu$_{2}$, CuO, and MgO.
The results, summarized in Table \ref{table:mol}, 
are in very good agreement with the experiment.
We also determined the equilibrium lattice constant $a$ and the bulk
modulus $B$ of bulk MgO.  
For these calculations cubic supercells with 64 atoms were
used with only $\Gamma$ k-sampling. 
Also in this case,
our results for $a$ and $B$ are close to the experimental values
(see Table \ref{table:bulk}). 

\section{Structure and energetics}
\label{struc_ener}

\subsection{Convergence tests}
\label{struc_ener.convtests}

Surfaces were modelled using a repeated slab geometry with periodic
boundary conditions parallel to the surface. 
A vacuum of thickness $d_V$ is introduced between slabs.
$d_V$ should be large enough to avoid spurious interactions between slabs. 
Each slab is composed of $N_L$ layers,
and surface supercells containing $(N_{at}/2)$ magnesium and $(N_{at}/2)$  
oxygen atoms per layer are used. 
In our structural optimizations, the lower surface
of the slab was kept fixed in a bulk-terminated configuration,
while other layers were fully relaxed. 

Tests with different values of $d_V$, $N_L$
and $N_{at}$ were performed for a few selected properties.
In Table \ref{table:convtests} the results of these tests for
the surface energy  $E_{surf}$, the binding energy  $E_{S-A}$
of a single adsorbed copper atom at the on-top oxygen site 
(see Eq. \ref{eqn:E_{S-A}}),
and the diffusion barrier $E_{diff}$ of a 
single adsorbed copper atom are reported.
$E_{diff}$ is defined as the difference between the Cu-surface
binding energies at the absolute minimum  (on-top of an oxygen surface
atom) and at the saddle point (hollow site, see below).
$E_{surf}$ is defined as :

\begin{equation}
E_{surf} = \frac{1}{2 A_{cell}}[E_{slab}^{tot} - N_L \frac{N_{at}}{2}
E_{bulk}]
\label{eqn:E_{surf}}
\end{equation}

\par\noindent
where $E_{slab}^{tot}$ is the total energy of the slab, $E_{bulk}$ is
the energy of an Mg-O pair in the bulk, and $A_{cell}$ is the
area of the supercell. The factor 2 accounts for the two exposed surfaces.
$E_{bulk}$ was obtained from a bulk calculation using a supercell with 64 atoms.

From Table \ref{table:convtests}, we can see that neither $E_{surf}$, 
$E_{S-A}$ nor $E_{diff}$ depend significantly on $d_V\;$. 
$E_{surf}$ depends weakly on $N_L$, whereas it depends 
strongly on $N_{at}$, i.e.  on the size of the supercell. 
This is in turn equivalent to
a dependence on the k-sampling of the surface Brillouin zone
(larger supercells corresponding to improved sampling).
The variation of $E_{diff}$ with $N_L$ and $N_{at}$ is very small.
The behavior of $E_{S-A}$ is slightly more complicated, but
we can see that also this quantity is very stable when large enough
supercells are used.

In Table \ref{table:convtests}  results of convergence tests
with respect to the plane-wave energy cut-off $E_{cut}^{w}$ 
are also reported. We note no substantial change bewteen 
the 16 Ry and the 20 Ry results for $E_{diff}$ and $E_{S-A}\;$
(in the case $N_{at}=18$).  Thus, we take $E_{cut}^{w}=16$ Ry
as our standard energy cut-off. 
On the basis of these tests, we
estimate that our calculated energy differences are accurate within
0.1 eV.

\subsection{Clean MgO surface}
\label{struc_ener.clean}

Before studying copper adsorption on MgO, we start by a characterization of
the clean surface.
The structure of the fully relaxed surface is very close to the ideal one,
as reported in other works\cite{ref:pg,ref:zlgh}. 
We find an inward surface relaxation of 1.2 \% of the bulk inter-layer spacing 
(0.02 \An \hspace{-1.5mm}).
This is accompanied by a rumpling of 1.5\%, 
with an outward displacement of the oxygens with respect to the
magnesium atoms. 
Our best estimate for the surface energy is 1.04 J/m$^2$ within the
LDA (see Table \ref{table:convtests}).
This value agrees with the results of 
previous theoretical work\cite{ref:pg,ref:t}.
The GGA leads to a remarkable decrease of the surface energy,
resulting in $E_{surf} = 0.86$ J/m$^2$.
A similar effect of GGA vs LDA has been found in recent work on
SnO$_2$ and TiO$_2$ surfaces\cite{ref:ghkgw}.

\subsection{Free Cu$_n$ clusters}
\label{struc_ener.free}

Cohesive energies and average distances for free clusters relevant to the
present study are reported in Table \ref{table:clus_res} (see also Fig.
\ref{fig:all_clusters}).
The calculations were performed placing the clusters in periodically repeated 
cells of sizes equal to the ones used for the adsorbed systems.
The optimized structures of these clusters were obtained
starting from the geometries of the adsorbed clusters on the surface.
Thus these structures generally correspond to a local minimum
in the potential energy surface.
The cohesive energy of the free clusters is calculated as:

$$D_{0} [n] = -E_{Cu_{n}} + n E_{Cu_{1}}\;,$$

\par\noindent
where $E_{Cu_n}$ is the total energy of the cluster with $n$ atoms.
The energy of the free copper atom, $E_{Cu_{1}}$, was
determined using the same pseudopotential and plane wave
cut-off employed for the cluster.
To account for polarization effects due to the unpaired electron 
in the atomic $4s^1$ state, we used the empirical correction 
$\Delta E_{LDA-LSD} = -0.18\; eV (n_{\uparrow} - n_{\downarrow})^2\;$
\cite{ref:jk}, where $n_{\uparrow}$ ($n_{\downarrow}$) is the number
of spin up (down) electrons. Our values for $D_{0}$ agree
very well with those of other theoretical works and experiments
(see Table \ref{table:clus_res}).
The general trend for $D_{0}$ is to increase as $n$ does.
Bonding distances are well reproduced.
The most stable geometry is the obtuse triangle for $n=3\;$ and the 
rhombus for $n=4\;$. 

\subsection{Supported clusters}
\label{struc_ener.supported}

In this section we present our results for the structure and energetics
of supported Cu$_{n}$ clusters of sizes $n = 1,2,3$ and $4\;$. 
The energetics will be characterized in terms of two quantities: 

\par\noindent
1. The cluster adsorption energy with respect to the substrate,
   $E_{S-A}$ :

\begin{equation}
E_{S-A} [n] = -[E_{Cu_{n}}^{ads} - E_{slab} - E_{Cu_{n}}^{free}]/n
\label{eqn:E_{S-A}}
\end{equation}

\par\noindent
where $E_{Cu_{n}}^{ads}$ is the total energy of the adsorbate-substrate 
system, $E_{slab}$ is the energy of the clean surface, 
and $E_{Cu_{n}}^{free}$ is the energy of the "free" cluster. 
As mentioned above,
this energy is obtained by optimizing the structure of the cluster 
starting from the geometry of the adsorbed cluster on the surface. 

\par\noindent
2. The intra-cluster binding energy $E_{A-A}$ :

\begin{equation}
E_{A-A} [n]= -[E_{Cu_{n}}^{ads} + (n-1) E_{slab} -
n\;E_{Cu_{1}}^{ads}]/n
\label{eqn:E_{A-A}}
\end{equation}

\par\noindent
where $E_{Cu_{1}}^{ads}$ is the total energy for the surface with
one adsorbed adatom. 

The structure of the cluster will be characterized in terms of
the average adsorbate-substrate distance ${\overline d}_{S-A}\;$,
the average copper-copper distance ${\overline d}_{A-A}\;$
and the angle $\alpha$ between an adsorbed copper atom, 
its supporting oxygen atom and the underlying magnesium atom.

All the results for the following sections
are summarized in Table \ref{table:ads_res}.
As the results for a single adsorbed copper atom show that the 
on-top oxygen site is strongly favored (see below), 
we choose to adsorb the cluster atoms at oxygen sites. 

\subsubsection{Cu$_{1}$}
\label{struc_ener.supported.1}

We considered three possible adsorption sites for a single copper atom 
on MgO(100): on top of the oxygen site, 
on top of the magnesium site and between two oxygen sites (hollow site, 
see Fig. \ref{fig:sites}).
The preferred site is on top of the oxygen 
atom (see Table \ref{table:ads_res}).
In the LDA we find an adsorption energy of 1.46 eV 
and an adsorption distance of 1.89 \An. 
The oxygen atom is attracted towards the Cu adatom and moves up 
by 0.1 \An with respect to the other oxygen atoms of the topmost layer. 
The (LDA) interpolated potential energy surface for the Cu adatom
is shown in Fig. \ref{fig:ener_surf}. The on-top-O site 
is the only minimum, magnesium is a maximum, 
while the hollow site is a saddle point
and thus corresponds to the transition state for diffusion.
The diffusion barrier is $E_{diff} = 0.45$ eV.
Assuming a simple Arrhenius-like expression for
the adatom attempt-to-jump frequency $\Gamma$, with a 
typical prefactor $\Gamma_{0} \sim 10$ THz,
we can estimate that at room temperature $\Gamma~\sim~\Gamma_{0}\;
\exp(-E_{diff}/kT)\;~\sim 280$~KHz, i.e. a residence time of
$3.6\times~10^{-6}$ s. 
This indicates that Cu adatom motion should be readily 
seen, for instance in Scanning Tunneling Microscopy (STM) experiments. 
Following Ref. \cite{ref:ss}, we can approximately define the
temperature for which the adatom diffusion becomes active (i.e. adatoms
jump at least once per second) as
$$T_d= \frac{E_{diff}}{k_{B}\;ln (4 D_0 / a^2)}\;,$$
where $D_0$ is the prefactor in the diffusion coefficent
and $a$ ($\sim 3$ \An) is the distance between neighboring adsorption sites.
When diffusion can be assimilated to a random walk (i.e. for $E_{diff}
\gg k_{B}\;T$), $\Gamma_{0}$ is related to $D_{0}$\cite{ref:bl1}. With
$\Gamma_{0} \sim 10$ THz, $D_0 \sim 2.3\times 10^{-3}$ cm$^{2}$/s, and
thus $T_d \sim 180\;$K (with an error margin of around 20 K,
corresponding to an error of one order of magnitude in $D_{0}$). 
This value of $T_d$ has been confirmed to be of the correct order 
by recent Electron Energy Loss Spectroscopy(EELS)
experiments\cite{ref:private}.

As the LDA is known to greatly overestimate binding energies,
(fully self-consistent) GGA calculations
(also including the optimization of the atomic structure)
have been carried out. These lead indeed to a reduction of the
adsorption energy of about 0.5 eV,
so that the resulting value is $E_{S-A} = 0.99$~eV. 
Within the GGA, also a significant increase of the Cu-O distance 
with respect to the LDA takes place (from 1.89 to 2.01 \An).
Note however that GGA calculations which employ the
LDA charge density and geometry yield 
a value of the binding energy (0.88 eV) which is very close
to that given by the full GGA calculations. 
For consistency with GGA results for larger clusters, the latter
(non self-consistent) value is reported in Table \ref{table:ads_res}.
From this Table we can also remark that the GGA diffusion barrier
is 0.45 eV, i.e. almost equal to that found within the LDA. 

Our LDA adsorption energy, adsorption distance 
and migration barrier agree well 
with previous LDA calculations\cite{ref:llp}, 
whereas there are several important discrepancies 
between our GGA results and the calculations 
of Pacchioni and R\"osch\cite{ref:pr},
who used cluster models along with
the B-LYP\cite{ref:b,ref:lyp} exchange-correlation functional.
At variance with most available (both experimental and theoretical)
results, these authors found that 
the Mg site is slightly more stable than the on-top O site, 
while the hollow site is only $\sim 0.1$ eV higher than the O site.
For the latter they obtained $E_{S-A} \sim 0.3$ eV 
and $d_{S-A} = 2.18$~\An, i.e. values which are much 
smaller and much larger than our calculated adsorption 
energy and distance respectively. 

There are two main factors which may contribute to 
the large discrepancies between our results and those of 
Pacchioni and R\"osch. One may be related to the convergence properties
of their results with respect to cluster size and embedding.
The other is the fact that the exchange-correlation functionals
used in the two calculations are different.

In an attempt to check the influence of the latter factor,
we performed a (fully self-consistent) calculation
using the Becke exchange-only functional
(again including the optimization of the atomic structure).
We stress that this calculation is only meant to test the dependence
of $E_{S-A}$ on the type of gradient corrected functional,
since it is widely accepted that the Becke-exchange-only
functional is generally less accurate than the GGA
(although for many molecules it yields dissociation 
energies which are in quite good agreement with experiment\cite{ref:b2}).
Using the Becke functional, we find that
the Cu-O adsorption distance becomes 2.09~\An 
(closer to the value of Ref. \onlinecite{ref:pr}),
while the binding energy is now extremely low,
0.19 eV {\em without} spin-polarization correction. 
This surprising result confirms a tendency to underbind
of the Becke exchange-only functional (see, e.g., 
Ref. \onlinecite{ref:shp}, where this tendency is found for
hydrogen-bonded systems). It also shows that 
Cu/MgO is a rather "difficult" case, with
a strong dependence of the adsorption energy on the
choice of the gradient-corrected functional. 
Although no well-defined experimental value of $E_{S-A}$
is available which may clearly identify the most appropriate
functional for the Cu/MgO system,
it is important to remark that recent EELS measurements\cite{ref:private}
suggest a value of the Cu diffusion barrier 
which is very close to our result of 0.45 eV, 
whereas a diffusion barrier of only $\sim 0.1$ eV
is inferred from the calculations of Ref. \onlinecite{ref:pr}.

\subsubsection{Cu$_{2}$}
\label{struc_ener.supported.2}

We considered three different starting geometries for a copper dimer 
on MgO(100): one with the two Cu's on nearest neighbor 
O atoms at distance of about 3 \An (A, see Fig. \ref{fig:sites_all}); 
the second (B) having the copper atoms on second neighbor oxygens 
at distance 4.2 \An, and a magnesium atom in between them;
the third (S, Fig. \ref{fig:sites_all}) with the copper dimer
perpendicular to the surface on top of an O atom.

After optimization, 
the dimer in configuration A is slightly stretched from 
its free geometry (from 2.18 \An in the free dimer to 2.25 \An) 
to satisfy bonding with the O atoms. $\alpha$ is 166\Deg. 
The oxygen atoms which support the adatoms move by 0.05 \An towards them.
The Cu-Cu cohesive energive is of order 0.8-0.9 eV (see Table 
\ref{table:ads_res}).
In configuration B, the two copper atoms cannot move as close as 
they would like to optimize their bond, because of the magnesium atom 
in between them. The Cu-Cu distance is now 2.34 \An, and the
Cu-Cu cohesive energy is decreased accordingly.
The two supporting oxygen atoms are displaced
from their ideal positions  towards the Cu's by 
rather large amounts, $\sim 0.14$ \An. 
Finally in configuration S, the copper dimer bond length
is almost identical to that of the free molecule, while 
the Cu-O distance is the same as that for the adsorbed copper adatom. 
The molecule is slightly tilted in the (110) plane, the O-Cu-Cu angle
being 175\Deg.

Within the LDA, the lowest energy structure is A
followed by S (for which the total energy is 0.22 eV higher 
than for A) and B (+0.33 eV w.r.t. A). Rather unexpectedly,
S is found to be the most stable structure within the GGA, 
even though the difference with A is quite small:
0.15 eV when the GGA is applied non-selfconsistently,
and 0.04 eV only when a fully selfconsistent GGA calculation
(including structural optimization) is performed.
This indicates a very delicate balance between intradimer cohesion
(dominant in the S configuration) and dimer-surface binding (dominant
for the A structure).

To investigate this issue further, we sampled the total energy surface 
as a function of the angle ($\theta$) between the dimer and the surface,
in a plane perpendicular to the surface and passing through 
two neighboring oxygen atoms (the (110) plane). We constrained the
value of $\theta$ and allowed all the other degrees of freedom to relax. 
The resulting S$\rightarrow$A energy barrier, occurring at 
$\theta \sim 20$\Deg, is $\Delta E^{dimer} = 0.13$ eV
(see Fig. \ref{fig:E_theta}). Frequent flips between the S and A structures
can be thus expected at room temperature. 
 
\subsubsection{Cu$_{3}$}
\label{struc_ener.supported.3}

The lowest energy structure for the free Cu$_3$ cluster 
is an obtuse triangle, corresponding to a Jahn-Teller distortion
of the equilateral triangle\cite{ref:ckrs,ref:j,ref:lb}. 
By contrast, we find that the linear arrangement 
on nearest neighbor O sites (configuration A, see Fig. \ref{fig:sites_all}) 
is the preferred adsorption geometry of the copper trimer. 
It is more bound to the surface and tighter 
(see the value for $\overline{d}_{S-A}$) 
than the triangular arrangement (configuration B).
This is due to the presence of a magnesium atom on one of the edges of  
the triangle, which prevents the two copper atoms 
from binding in an optimal way. 
In the triangular arrangement, all three copper atoms are 
tilted towards each other: $\alpha$ is 166\Deg$\;$ for the central
atom
and 162\Deg$\;$ for the other two. The oxygen atoms move only
slightly out
of their equilibrium positions.
In configuration A, the copper atoms at the two ends of the line 
are tilted towards the middle adatom
to achieve optimal Cu-Cu distances ($\alpha = 159$\Deg).
Oxygen atoms which support copper atoms at the end of the line 
are attracted to them and move up by 0.08 \An 
while no difference can be seen for the middle atom. 

\subsubsection{Cu$_{4}$}
\label{struc_ener.supported.4}

To determine the most stable arrangement of adsorbed Cu$_{4}\;$,
two different starting geometries were considered: a linear
configuration on nearest neighbour O atoms (A), 
and a square geometry with the four Cu's on top of nearest
neighbor surface O atoms and a Mg atom at the center of the square (B),
similar to one considered by Pacchioni and R\"osch\cite{ref:pr} (see
Fig. \ref{fig:sites_all}).

We find that upon relaxation the square changes its shape 
to a rhombus, which, as in the case of the free Cu$_4$
cluster\cite{ref:mpc2,ref:ckrs,ref:j}, becomes the lowest 
energy structure for the adsorbed tetramer.
However, the adsorbed rhombus is not perfectly planar : 
the Cu atoms at the end of the short diagonal 
attract their supporting oxygens by 0.11 \An ($d_{S-A}$ = 2.11 \An) 
with $\alpha = 154^{o}$, while the other two Cu atoms, staying almost 
on top of their oxygens, lower them with respect to other surface 
atoms by 0.07~\An ($d_{S-A}$ = 2.17 \An \hspace{-1mm}) with $\alpha = 175^{o}$
(see Fig. \ref{fig:sites_all}).
The adsorption energy per atom of this structure 
is very low (see Table \ref{table:ads_res}), and 
close to that found by Pacchioni and R\"osch\cite{ref:pr} 
for Cu$_4$ in a square geometry.

Upon relaxation, the linear Cu$_4$ cluster
($\sim 0.7$ eV higher in energy than the rhombus)  
splits into two dimers of length 2.28 \An, separated by
a distance of 3.28 \An (see Fig. \ref{fig:sites_all}).
However, these dimers are not totally independant,
since they show a mirror plane symmetry with respect to the 
center of the line: $\alpha = 160^{o}$ 
for the extreme atoms  and $\alpha = 174^{o}$ for the inner atoms. 
The oxygen atoms supporting the extremity of the line move up by 0.06 \An, 
attracted by the adsorbate atoms, while the other two lower 
their positions by 0.02 \An.

\subsubsection{Trends}
\label{struc_ener.supported.trends}

Competition between adsorption on the surface and cohesion
among the atoms in the cluster is an important feature of the Cu/MgO system. 
In the $n=2$ case, the dimer (configurations A and B) is stretched
from its equilibrium distance in the gas-phase to take into account 
interaction with the surface.  The balance between intra-dimer
and surface-dimer bonding is delicate: in fact full GGA calculations
yield very close energies for the A and S (standing) geometries.
In the $n=3$ case, the linear structure is preferred to the triangular
arrangement, which is more stable in the gas-phase, because it
allows a better bonding to the substrate.
For $n=4$, instead, the rhombus, which is by far the most stable
stable isomer in the gas-phase, is also preferred for the adsorbed cluster,
in spite of its low adsorption energy.
In the $n=4$ linear case, the cluster prefers
to split into two separate dimers in order to optimize its structure
and its bonding with the supporting oxygens.

\indent In Fig. \ref{fig:E_bind} our results for 
the surface-adsorbate ($E_{S-A}$, c) and d)) and adsorbate-adsorbate 
intra-cluster ($E_{A-A}$, a) and b)) binding energies are summarized.
We can remark that the effect of gradient corrections is more
important on  $E_{S-A}$ than on $E_{A-A}$,  
and the absolute difference between LDA and GGA results
for $E_{S-A}$ tends to reduce with $n$.
With increasing $n$, the
adsorption energy per atom $E_{S-A}$ decreases, while $E_{A-A}$ 
first decreases (from $n=2$ to $n=3$) and then increases.
Calculations for $n>4$ clusters\cite{ref:m_else}  
confirm the latter trend. This is an indication that,
provided deposition on the surface is sufficiently ``soft'', 
Cu$_n$ clusters will tend 
not to wet the surface but to retain their cluster-like character,
particularly in the case of large clusters. 
Our results also show that larger (adsorbed) aggregates are
energetically more stable than smaller ones.
Thus when Cu atoms are deposited on the surface to grow a film,
at high enough temperature formation of large aggregates by
coalescence of small diffusing clusters, particularly monomers,
could be observed.\\
\indent Calculations for $n>4$\cite{ref:m_else} also show that
3D geometries are energetically preferred to 2D ones. In particular
for $n=5$ the lowest energy configuration is a square pyramid
($E_{S-A} = 0.43$ eV, $E_{A-A} = 1.08$ eV, within GGA). 
The total energy of this configuration is found to be 0.95~eV
lower than that of the planar trapezoidal geometry.

\section{Electronic properties}
\label{elec}

\subsection{Clean MgO surface}
\label{elec.clean}

The (LDA) density of occupied states (DOS) -- obtained by artificially
broadening the one-electron eigenvalues with gaussians
of width 0.15 eV -- is displayed in Fig. \ref{fig:1-2-3-4_ads}.
This shows oxygen 2$s$ (region I) and 2$p$ bands (regions II and III) 
separated into smaller bands.

Since only occupied states are included, magnesium states should not be seen 
if one considers the MgO(100) surface to be completely ionic and 
hence the magnesium valence states to be empty. 
Recent calculations\cite{ref:sa} have nonetheless shown that some
charge stays on the magnesium instead of going to the oxygens. This is
also the case in our study when one looks at the projected DOS of
Fig. \ref{fig:proj_dos}. 

The overall bandwidth (difference between the top of the O$2p$
and the bottom of the O$2s$ bands) is 17.0 eV, while
the width of the O$2p$ band is 4.0 eV.
Corresponding experimental values by R\"ossler {\it et al.}\cite{ref:rw}
are 20.0 and 5-6 eV respectively. 
We also calculated the bandgap between occupied and 
empty states and found $E_{gap}=3.0\;$ eV,
against an experimental value of 7.8 eV\cite{ref:wfw} for the
bulk. The surface band gap has been measured and calculated to be of
around 4.5 eV\cite{ref:tvs} and 5.0 eV\cite{ref:lw} respectively.
The strong underestimate of the gap is a well-known problem of density
functional theory.

\subsection{Free Cu$_n$ clusters}
\label{elec.free}

In order to help understanding the bonding mechanism 
between the Cu$_n$ clusters and the MgO(100) surface, 
in Fig. \ref{fig:1-2-3-4} we show the DOS of the free 
clusters obtained by optimizing the adsorbed structures. 
Similar DOS for Cu$_n$ up to $n=5$ have been discussed 
previously by Jackson\cite{ref:j}.
Our monomer's DOS shows the small separation between the 4s state (at $\sim$
-4.5 eV) and the 3d state (at $\sim$ -5.5 eV). 
For the dimer, this split-off state on the high energy side of the
3$d$ band is missing. This indicates that the dimer is a closed-shell system
\cite{ref:ckrs}. This characteristics holds also for
the Cu$_3$ and Cu$_4$ {\sl linear} clusters, whereas
a split-off state on the high-energy side of the DOS is present
for the {\sl triangular} Cu$_3$ and for the {\sl rhombus}.
The character of the high-energy state is mainly s-like, 
as found by Jackson\cite{ref:j} and Massobrio {\em et al.}\cite{ref:mpc2}.
In analogy with the work of Li {\em et al.}\cite{ref:llp},
we shall call the split-off state 4$s^*$. 

\subsection{Supported clusters on MgO}
\label{elec.supported}

In this section we try to elucidate the bonding
between the Cu$_n$ clusters and the MgO(001) surface.
To this end we shall examine the DOS for the adsorbed clusters
as well as the charge density (for the case $n=1$).

\subsubsection{Cu$_{1}$}
\label{elec.supported.1}
The upper panel of Fig. \ref{fig:1-2-3-4_ads} shows the (LDA) DOS
for a single copper atom at the oxygen and magnesium sites of the
MgO(100) surface. 
A split-off state appears on the top of the bands (region IV).
This state is well inside the bandgap of the clean MgO(100) surface,
and causes its reduction from a value of 3.0 eV for the clean
surface to 1.5 eV for the on-top O site.
In the magnesium site case, the feature corresponding to the 4s$^*$
state is smaller.  
For the O site, prominent states are present 
in region II.
These states are responsible for the Cu-surface bonding 
as shown by Li {\em et al.} and below. 

In Fig. \ref{fig:proj_dos} we present the $s,p,d$ decomposition
of the DOS for copper at the stable adsorption on-top O site.
This decomposition was performed by projecting the plane wave basis
onto spherical waves centered on the Cu, O, and Mg atoms.
Inspection of Fig. \ref{fig:proj_dos}  shows that
the split-off state originates as expected from the Cu$4s$ state. 
Here no contribution from the $d$ states is present.
Regions II and III consist mainly O$2p$ 
states and Cu$3d$ states, the contribution from the Cu$4s$ state 
being small but not negligible especially near the band edges.
Region I contains almost exclusively O$2s$ states.
Magnesium states appear in regions II and III, 
indicating that the ionicity of the surface might not be complete. 

To further investigate the character of the bonding,
in Fig. \ref{fig:cd_all} we present plots of the (LDA) charge density  
for the various spectral regions of the DOS. 
Here, the contours of the integrated charge density 
are shown in a plane perpendicular 
to the surface and containing only O atoms.
For technical reasons, the part of the wave functions 
localized at the cores is missing from the plots. 
In region I, the charge density has the character
of $s$ states centered on the oxygens. The charge is distorted towards
the overlying copper atom. In region II, 
it clearly indicates the presence of a bond through overlap 
between oxygen $2p$ and copper $3d/4s$ states.
Region III shows little overlap between oxygen and copper orbitals.
In region IV the split-off state has anti-bonding character. 
In agreement with the LDA calculations by Li {\em et al.}\cite{ref:llp},
the above analysis of the charge density 
indicates that the bonding between a copper adatom and MgO(001)
is due to a mixing of the Cu $3d/4s$ states 
and the oxygen 2$p$ band. The bonding part lies
in region II, while the antibonding counterpart is in region IV.

The fact that the states
in region II are those mostly
responsible for the Cu-surface bonding is further confirmed by 
Fig. \ref{fig:dos_comparison} which compares the DOS for the $n=1$
case calculated using the LDA, the GGA, and the Becke exchange-only
functional. There is clearly a relationship between
$E_{S-A}$ and the importance of these bonding states. In fact, the 
LDA which yields the highest adsorption energy also shows the most
prominent features in the DOS, while in the ``Becke-only'' DOS these
peaks are missing.

\subsubsection{Cu$_{2}$,Cu$_{3}$,Cu$_{4}$}
\label{elec.supported.2}

The DOS for the adsorbed dimer shows
few closely spaced states near the top of the O-$2p$ band 
(see Fig. \ref{fig:1-2-3-4_ads}). 
Furthermore, for the standing dimer (S)
the density of states
in region II 
is very similar to that of a single copper atom at the on-top-O site.
This indicates that the intra-molecular bond is very little
perturbed in this case.
The closed shell nature of the dimer\cite{ref:ckrs} 
can explain why it prefers 
to adsorb vertically on the surface (at the GGA level).
This can also explain why the adsorption energy drops by $\sim 0.4$ eV 
from $n=1$ to $n=2\;$ (see Fig. \ref{fig:E_bind}, c) and d)). 
At variance with the free-cluster case, 
for $n=3$ the linear arrangement 
(configuration A, see Fig. \ref{fig:sites_all}) 
is preferred to the triangular one (configuration B). 
As mentioned previously, this can be attributed to the fact that for 
configuration B an Mg atom is present in between a pair of Cu atoms,
thus hindering their bonding.
From an electronic point of view, the bonding states
in region II are much more prominent for the linear case 
than for the triangular one. 
We can also remark the presence of two separate 4s$^*$ peaks
for the linear arrangement. 
For $n=4$, the linear tetramer splits into two dimers 
(configuration A of Fig. \ref{fig:sites_all}) 
and displays a DOS similar in shape to that of the dimer.
A similarity with the linear trimer case can be also noted. 
For the rhombus a well defined band above the top of the
O-$2p$ band is present, while the bonding states 
in region II are less prominent than for the
linear arrangement. Indeed the substrate-adsorbate
binding energy $E_{S-A}$ for the rhombus is lower
than for the linear tetramer (see Table \ref{table:ads_res}).
Viceversa, $E_{A-A}$ for the rhombus is substantially larger
than for the linear case. This is consistent with the fact that
for free Cu$_4$ clusters the rhombus is much more stable
than all other isomers \cite{ref:ckrs}.  

\subsubsection{Trends}
\label{elec.supported.trends}

When $n$ increases, the gap between the 4$s^*$ state 
and the top of the nearest occupied band is progressively filled; 
the formation of an independant band is
clearly shown in Fig. \ref{fig:1-2-3-4_ads}. 
Another interesting feature is the reduction of the DOS feature
corresponding to the bonding states
in region II.
This is a confirmation of the fact that 
with increasing their size the adsorbed
copper clusters reduce their ability to bond to the surface. 
Intracluster bonding becomes dominant and the copper agregates 
thus tend to retain their gas-phase structure. 

\section{Conclusion}
\label{conc}

This paper reports a first principles study of copper microclusters adsorbed
on an MgO(100) surface. At variance with previous calculations
where cluster models of rather small size 
were generally used to represent the surface, 
in our study we use a slab model and periodically repeated 
supercells of large size (up to 36 atoms per layer).  
Moreover full relaxation, without symmetry constraints,
of all atoms in the cluster and in the surface layer
is allowed for.

In agreement with previous studies,
our results show that the preferred site of adsorption
for a single Cu adatom on defect-free MgO(100) is the oxygen site. 
The calculated binding energy and distance are found to 
depend significantly on the type of gradient corrections used.
According to our best estimate, obtained using the GGA,
the binding energy is of about 1 eV, suggesting a weak covalent bond. 
This is confirmed by the character of the electronic charge density,
which shows the mixing of copper $3d$ and $4s$ states 
with oxygen $2p$ orbitals. The Cu hopping barrier is $\sim 0.45$ eV
(largely independent of the functional), suggesting that
significant adatom diffusion should be seen starting from
$\sim 180$~K. 

When increasing the size of the clusters, we find that 
the intra-cluster binding energy tends to dominate 
over the adsorption energy, larger (adsorbed) aggregates 
being energetically more stable than smaller ones.
These results suggest that the clusters 
which are ``softly'' deposited on the surface at low temperatures
will tend to keep their gas-phase structure (see, e.g., Cu$_2$ and Cu$_4$) 
while optimizing their bonding with the substrate.  
Viceversa, when Cu atoms are deposited on the surface 
to grow a Cu film on MgO(100), formation of large aggregates by
coalescence of small diffusing clusters (particulary monomers) 
should be observed at high enough temperature. 
Results for $n>4$ (to be presented elsewhere) show that bigger copper
clusters tend to form three-dimensional structures instead of planar
ones.

\section{Acknowledgements}
\label{aknow}
We are very grateful to A. Pasquarello for 
many helpful discussions and assistance with the 
Car-Parrinello-Vanderbilt LDA/GGA plane waves code. 
We are also pleased to thank M.-H. Schaffner, F.
Patthey, Prof. W. Schneider, and colleagues at IRRMA for useful
discussions. The calculations have been performed on NEC SX-3 and SX-4
machines at the CSCS in Manno, Switzerland, and on a CRAY Y-MP
computer at the EPFL in Lausanne, Switzerland. One of us (V.M.)
acknowledges the Swiss National Science Foundation for support under
Grants No. 20-39528.93 and 20-49486.96.

\bibliography{citations}

\begin{center}
\begin{table}
\caption{Calculated (LDA) bond distances and stretching frequencies of
a few selected molecules.}
\vspace{.5cm}
\begin{center}
\begin{tabular}{ccccc}
Molecule&\multicolumn{2}{c}{$d$ [\AA]}&\multicolumn{2}{c}{$\nu$ [eV]} \\
&LDA&Exp.\cite{ref:rs,ref:hh}&LDA&Exp.\cite{ref:rs,ref:hh}\\ \hline
Cu$_{2}$&2.18&2.22&0.034&0.033\\ 
CuO&1.69&1.72&0.087&0.079\\ 
MgO&1.77&1.75&0.103&0.097\\ 
\end{tabular}
\end{center}
\label{table:mol}
\end{table}

\begin{table}
\caption{Calculated structural parameters for bulk MgO. $a$ is the
lattice constant, $B$ is the bulk modulus, $V_{0}$ and $B'$ are
parameters from the Murnhagen equation of state.} 
\begin{center}
\begin{tabular}{lclcccc}
\multicolumn{3}{c}{}&$a$ [\AA] & $B$ [Mbar] & $V_{0}$ [\AA$^{3}$]&$B'$
\\ \hline 
This work&(LDA)&Murnhagen fit& 4.25&1.63&68.74&3.70\\
&(LDA)&Polynomial fit&4.25&1.57&68.63&-\\ \hline 
Theory&(Hartree-Fock)&Ref. \onlinecite{ref:cdpr}&4.20 &1.86&66.18&3.53\\ \hline
Exp.&-&Ref. \onlinecite{ref:ga,ref:aa,ref:sps,ref:dlcp}&4.21&1.55-1.62&66.60&-\\
\end{tabular}
\label{table:bulk}
\end{center}
\end{table}

\begin{table}
\caption{Convergence tests for the surface energy ($E_{surf}$), the
Cu-surface binding energy ($E_{S-A}$) and the Cu diffusion barrier
($E_{diff}$). Calculations are within the LDA. $N_{L}$ is the number
of layers, $N_{at}$ is the number 
of atoms per layer, $d_{v}$ is the width of the vacuum. $E_{cut}^{w}$
is the energy cut-off for the smooth part of the wave-functions.}
\begin{center}
\begin{tabular}{ccccccc}
$N_{L}$&$N_{at}$&$E_{cut}^{w}$ [Ry]&d$_v$ [\AA]&$E_{surf}$
[J/m$^2$]&$E_{S-A}$ [eV]&$E_{diff}$ [eV]\\ \hline
2&8&16&7.4&1.59&1.71&-\\
3&8&16&7.8&1.60&1.29&0.53\\
3&8&16&19.8&1.77&1.25&0.51\\
3&8&20&13.8&-&1.50&0.53\\
2&18&16&12.8&1.19&1.46&0.45\\
2&18&20&12.8&-&1.50&-\\
3&18&16&10.7&1.19&1.48&-\\
2&36&16&12.8&1.04&-&-\\
\end{tabular}
\end{center}
\label{table:convtests}
\end{table}

\begin{table}
\begin{minipage}{17cm}
\caption{Cohesive energy per atom ($D_{0}/n$) and average Cu-Cu
distance for the free clusters shown in Fig.
\protect\ref{fig:all_clusters}. The values for the cohesive energy have
been obtained within the GGA(LDA). Spin-polarization effects are
included in the works of Jackson and Calaminici
{\em et al.}, but not in that of Lammers {\em et al.}.}
\vspace{.5cm}
\begin{center}
\begin{tabular}{cllcc}
n&Geometry&Ref.&$D_{0}/n$ [eV]&${\overline d}$ [\An]\\ \hline
2&Line&This work&1.13 (1.33)&2.18\\
2&Line&Jackson \cite{ref:j}&1.08 (1.36)&2.18\\
2&Line&Calaminici {\em et al.}\cite{ref:ckrs}&1.13 (1.30)&2.20\\
2&Line&Exp\cite{ref:hh,ref:rv}&1.04&2.22\\
3&Obtuse triangle ($C_{2v}$)&This work&1.13 (1.43)&2.31\\
3&Obtuse triangle ($C_{2v}$)&Jackson \cite{ref:j}&1.16 (1.52)&2.27\\
3&Obtuse triangle ($C_{2v}$)&Calaminici {\em et
al.}\cite{ref:ckrs}&1.12 (1.34)&2.34\\
3&Obtuse triangle ($C_{2v}$)&Exp.\cite{ref:wv}&1.02&-\\
3&Line&This work&1.10 (1.36)&2.23\\
3&Line&Lammers {\em et al.}\cite{ref:lb}&0.57&-\\
3&Line&Calaminici {\em et al.}\cite{ref:ckrs}&1.18 (1.48)&2.27\\
4&Rhombus ($D_{2h})$&This work&1.53 (1.89)&2.32\\
4&Rhombus ($D_{2h})$&Jackson \cite{ref:j}&1.52 (1.95)&2.30\\
4&Rhombus ($D_{2h})$&Lammers {\em et al.} \cite{ref:lb}&1.04&-\\
4&Rhombus ($D_{2h})$&Calaminici {\em et al.}\cite{ref:ckrs}&1.59 (1.90)&2.36\\
4&Line&This work&1.48 (1.74)&2.24\\
\end{tabular}
\end{center}
\label{table:clus_res}
\end{minipage}
\end{table}
\vspace{.5cm}

\begin{table}
\caption{Calculated energies (in eV) and structures (distances in
\An \hspace{-1mm}) for adsorbed copper clusters shown in Fig.
\protect\ref{fig:sites_all}. All GGA results in
this Table are from non-selfconsistent calculations which employ the
LDA charge density and geometry.}
\begin{center}
\begin{tabular}{cccccccc}
&&\multicolumn{2}{c}{LDA}&\multicolumn{2}{c}{GGA}&LDA&LDA\\
n&Geometry&$E_{S-A}$&$E_{A-A}$ &$E_{S-A}$&$E_{A-A}$&${\overline
d}_{S-A}$&${\overline d}_{A-A}$\\ \hline
1&On top of O		&1.46&-   &0.88&-   &1.89&-\\
1&On top of Mg		&0.45&-   &0.18&-   &2.51&-\\
1&Hollow		&1.01&-   &0.43&-   &2.07&-\\ \hline
2&O-Cu-Cu-O (A)		&1.07&0.93&0.58&0.83&1.99&2.25\\ 
2&O-Cu-Mg-Cu-O (B)	&0.90&0.77&0.47&0.72&2.11&2.34\\
2&Standing on O	(S)	&0.96&0.82&0.71&0.96&1.87&2.20\\ \hline
3&Line on O (A)		&0.96&0.86&0.52&0.74&2.02&2.30\\
3&Triangle on O	(B)	&0.85&0.82&0.45&0.70&2.06&2.51\\ \hline
4&Rhombus on O (A)	&0.65&1.08&0.28&0.93&2.14&2.32\\
4&Line on O (B)		&0.80&0.89&0.39&0.81&2.02&2.59\\
\end{tabular}
\end{center}
\label{table:ads_res}
\end{table}
\end{center}

\begin{figure}[p]
\vspace{1cm}
\centerline{\psfig{figure=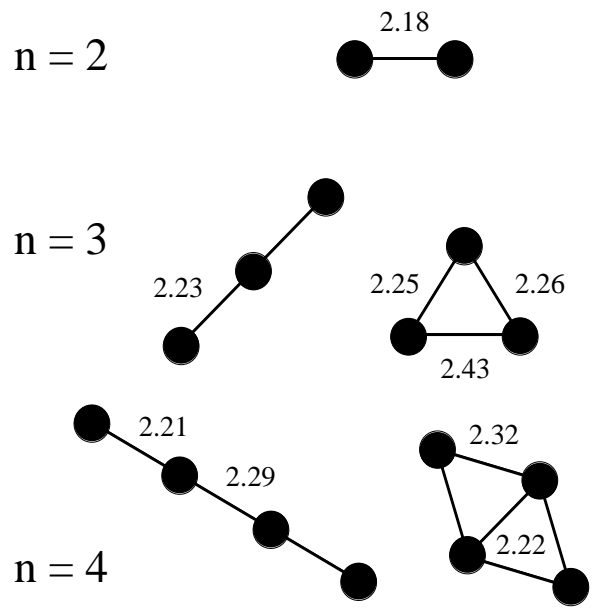,height=8cm}}
\vspace*{5cm}
\caption{Musolino {\em et al.}, Journal of Chemical Physics}
\label{fig:all_clusters}
\end{figure}

\capt{\protect\ref{fig:all_clusters}}{Free clusters obtained in LDA by
relaxing their adsorbed geometries. Distances are in \An.}

\begin{figure}[p]
\centerline{\psfig{figure=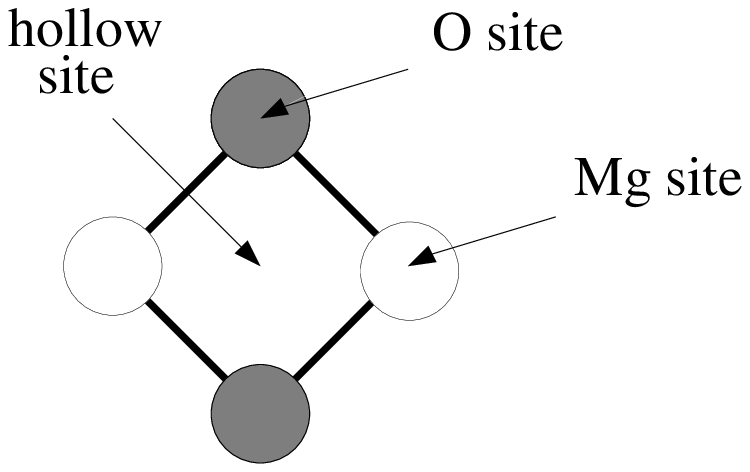,height=4cm}}
\vspace*{5cm}
\caption{Musolino {\em et al.}, Journal of Chemical Physics}
\label{fig:sites}
\end{figure}

\capt{\protect\ref{fig:sites}}{Copper adsorption sites on MgO(100).}

\begin{figure}[p]
\centerline{\psfig{figure=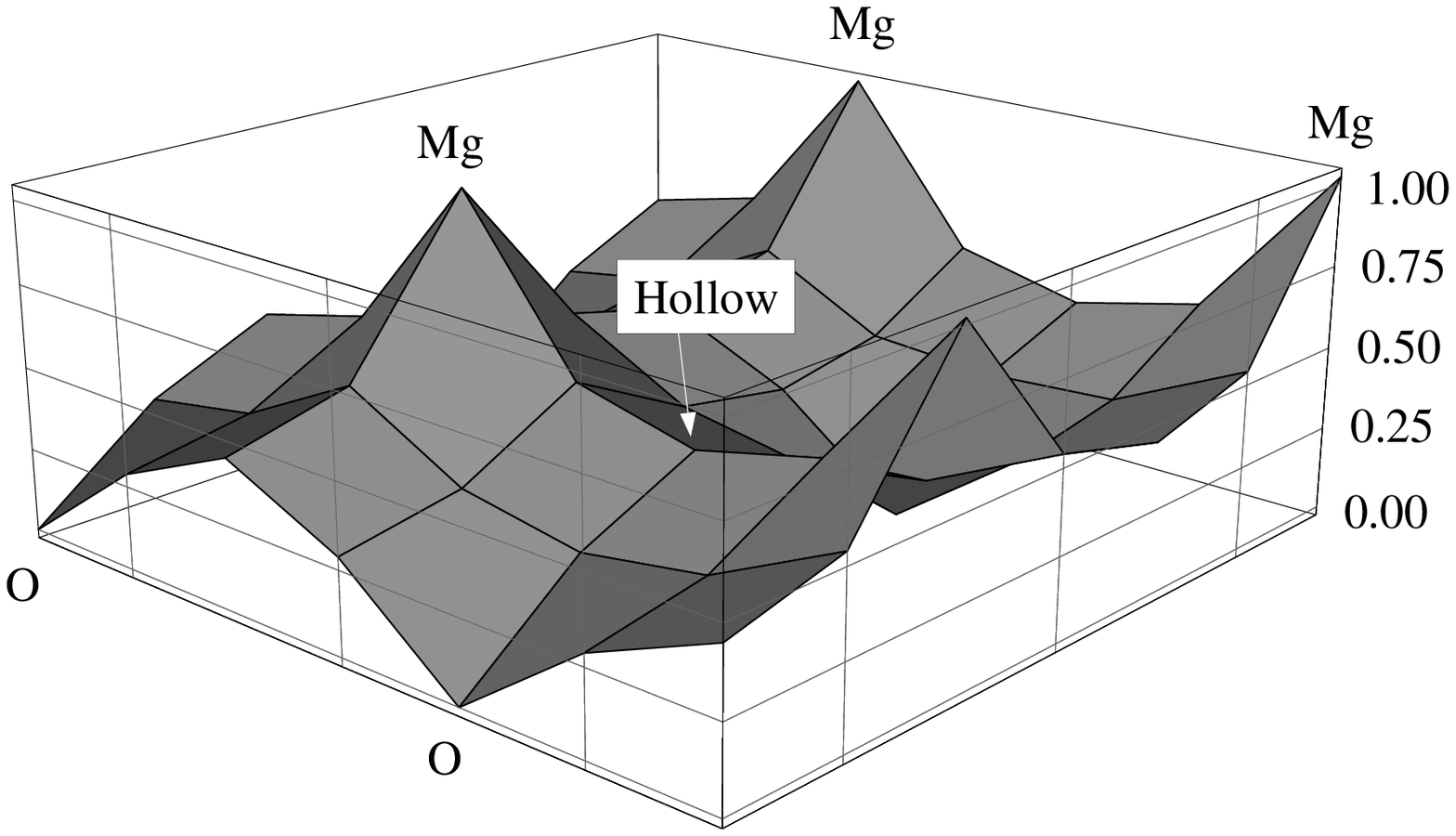,height=8cm}}
\vspace*{5cm}
\caption{Musolino {\em et al.}, Journal of Chemical Physics}
\label{fig:ener_surf}
\end{figure}

\capt{\protect\ref{fig:ener_surf}}{LDA potential energy surface for a
Cu adatom on MgO(100). Energy units are in eV and taken with respect
to the lowest lying state, i.e. the copper on top of the oxygen site.}

\begin{figure}[p]
\centerline{\psfig{figure=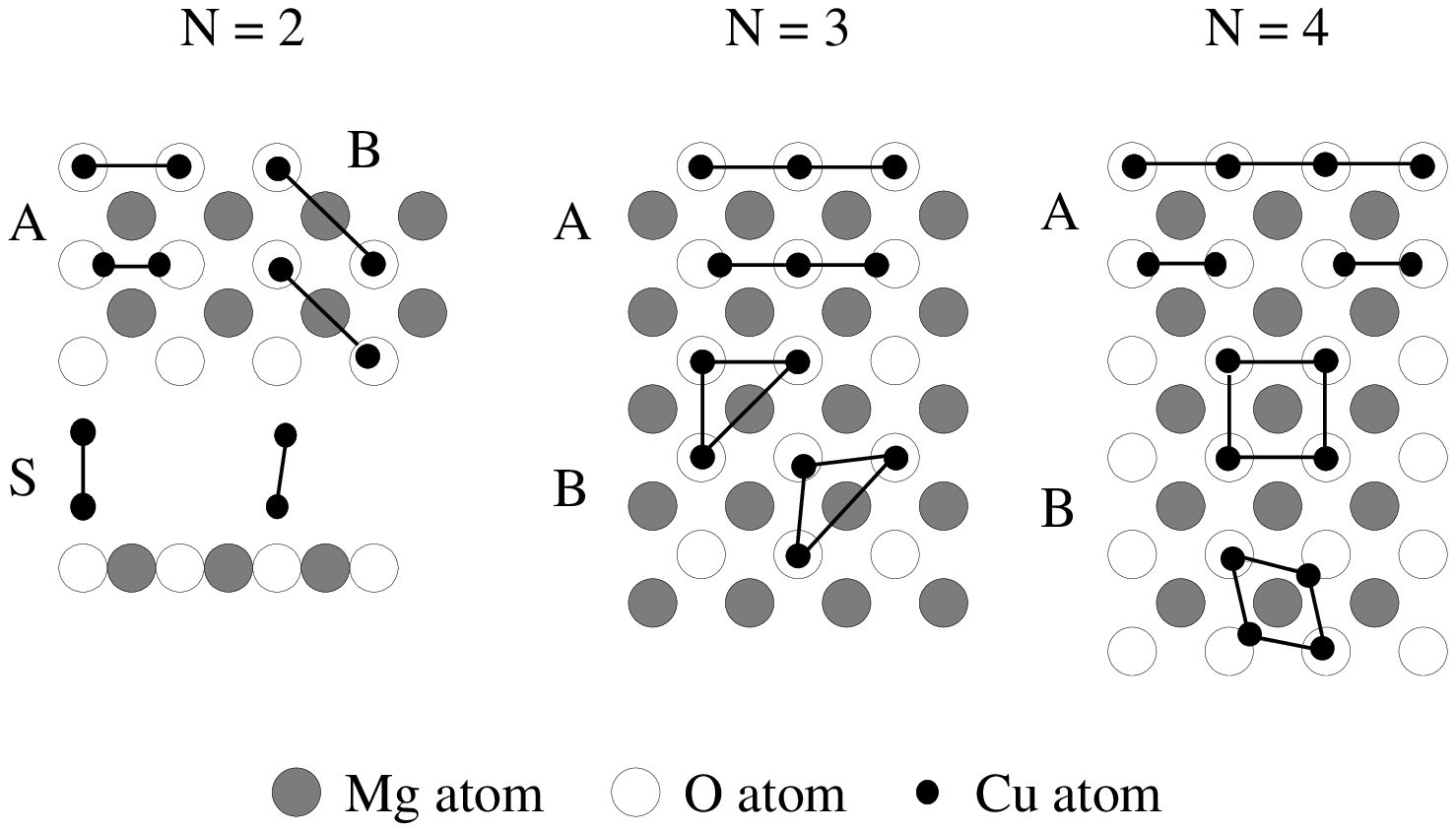,height=8cm}}
\vspace*{5cm}
\caption{Musolino {\em et al.}, Journal of Chemical Physics}
\label{fig:sites_all}
\end{figure}

\capt{\protect\ref{fig:sites_all}}{Geometries of the supported
Cu$_{n}$ clusters. For each case, we show both the starting geometry
(top or left) and the optimized structure (bottom or right).}

\begin{figure}[p]
\centerline{\psfig{figure=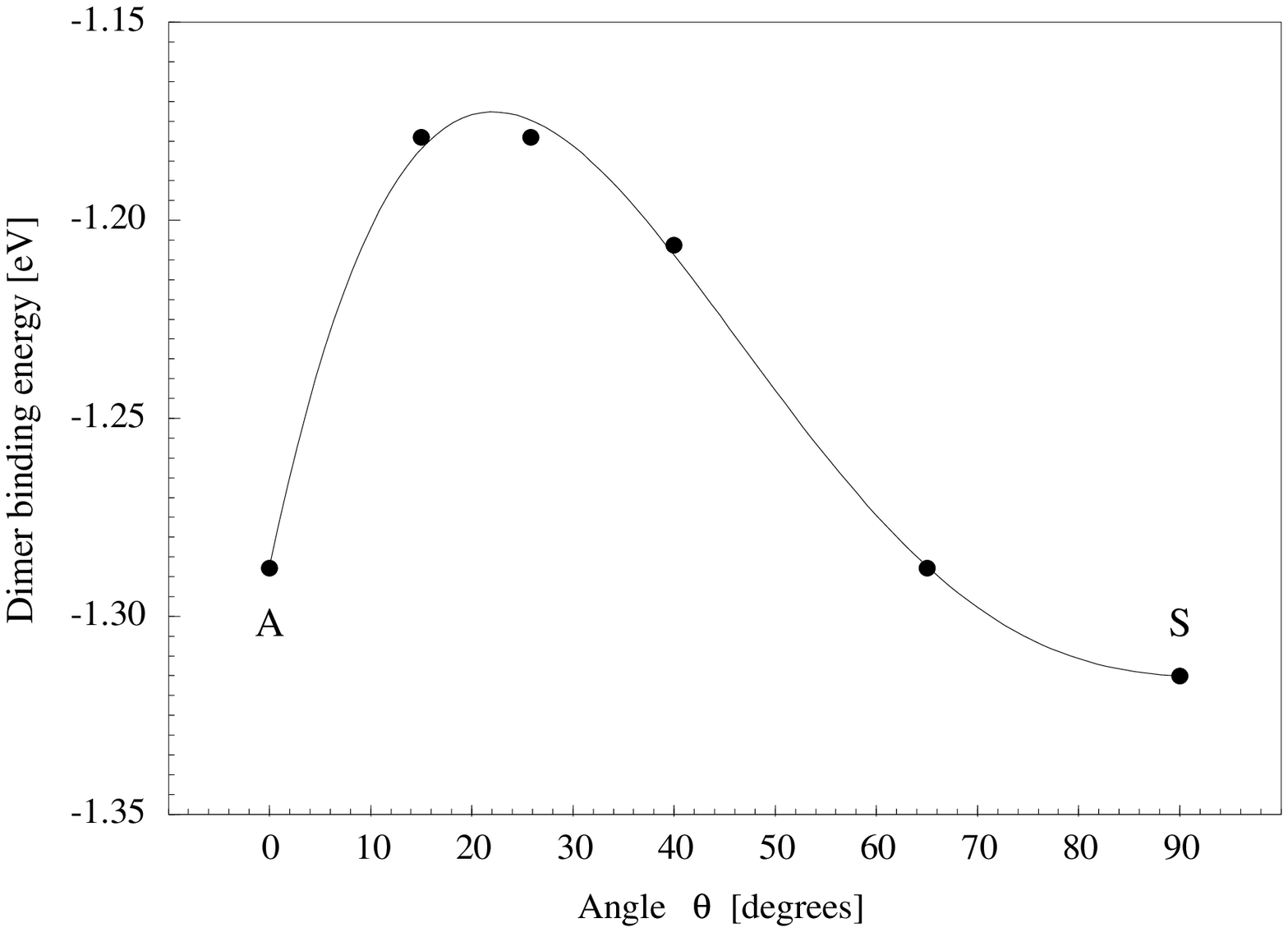,height=12cm}}
\vspace*{5cm}
\caption{Musolino {\em et al.}, Journal of Chemical Physics}
\label{fig:E_theta}
\end{figure}

\capt{\protect\ref{fig:E_theta}}{Dimer binding energy as a function of
the angle $\theta$ (see text for definition) as computed in the
GGA. The line is only meant as a guide for the eye.}

\begin{figure}[p]
\centerline{\psfig{figure=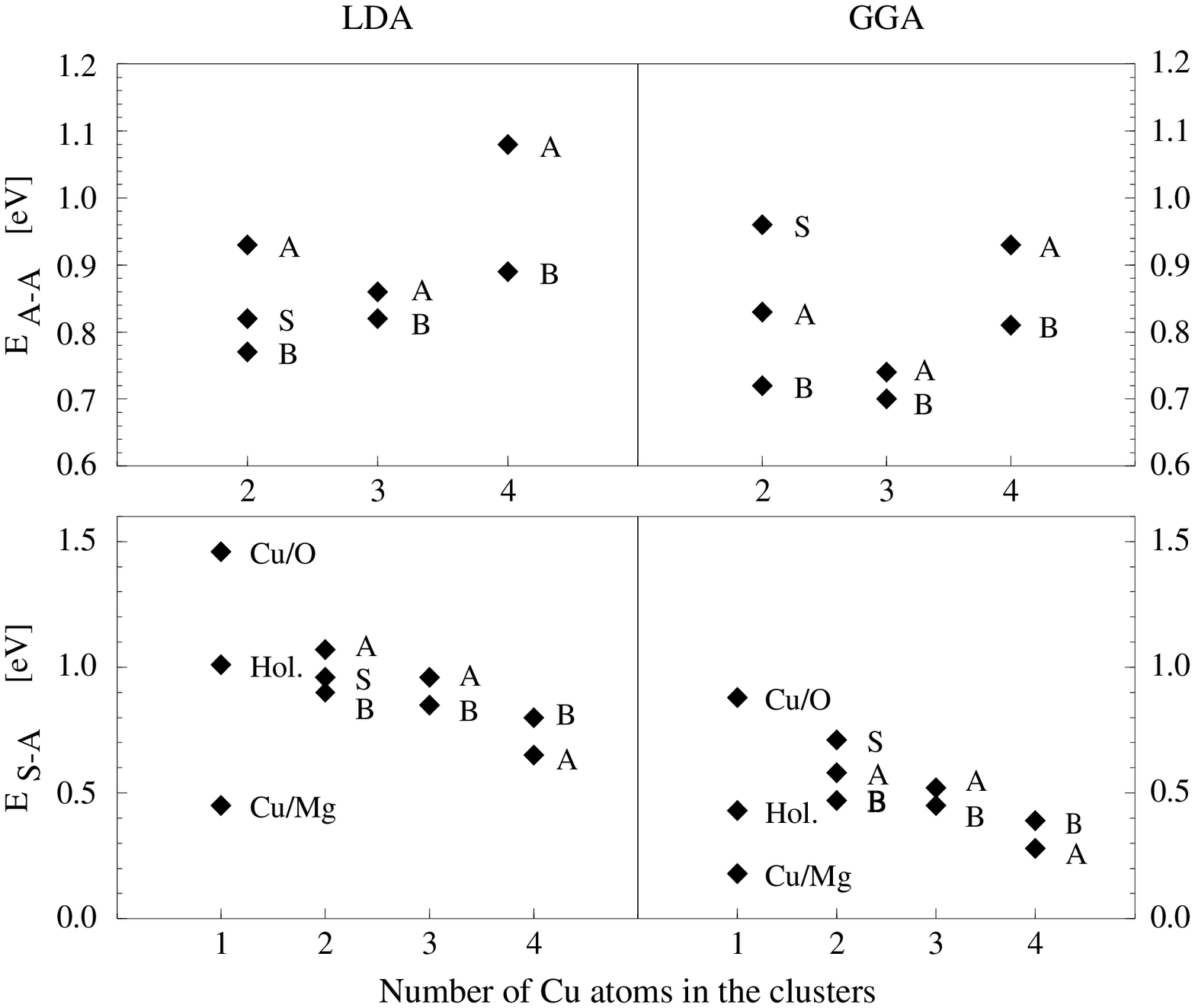,height=16cm}}
\vspace*{5cm}
\caption{Musolino {\em et al.}, Journal of Chemical Physics}
\label{fig:E_bind}
\end{figure}

\capt{\protect\ref{fig:E_bind}}{Binding energies $E_{A-A}$
and $E_{S-A}$ of copper clusters on MgO(100). Legends refer to the
geometries of Fig. \protect\ref{fig:sites_all}. GGA results are
obtained through a non-selfconsistent method (see text for details).}

\begin{figure}[p]
\centerline{\psfig{figure=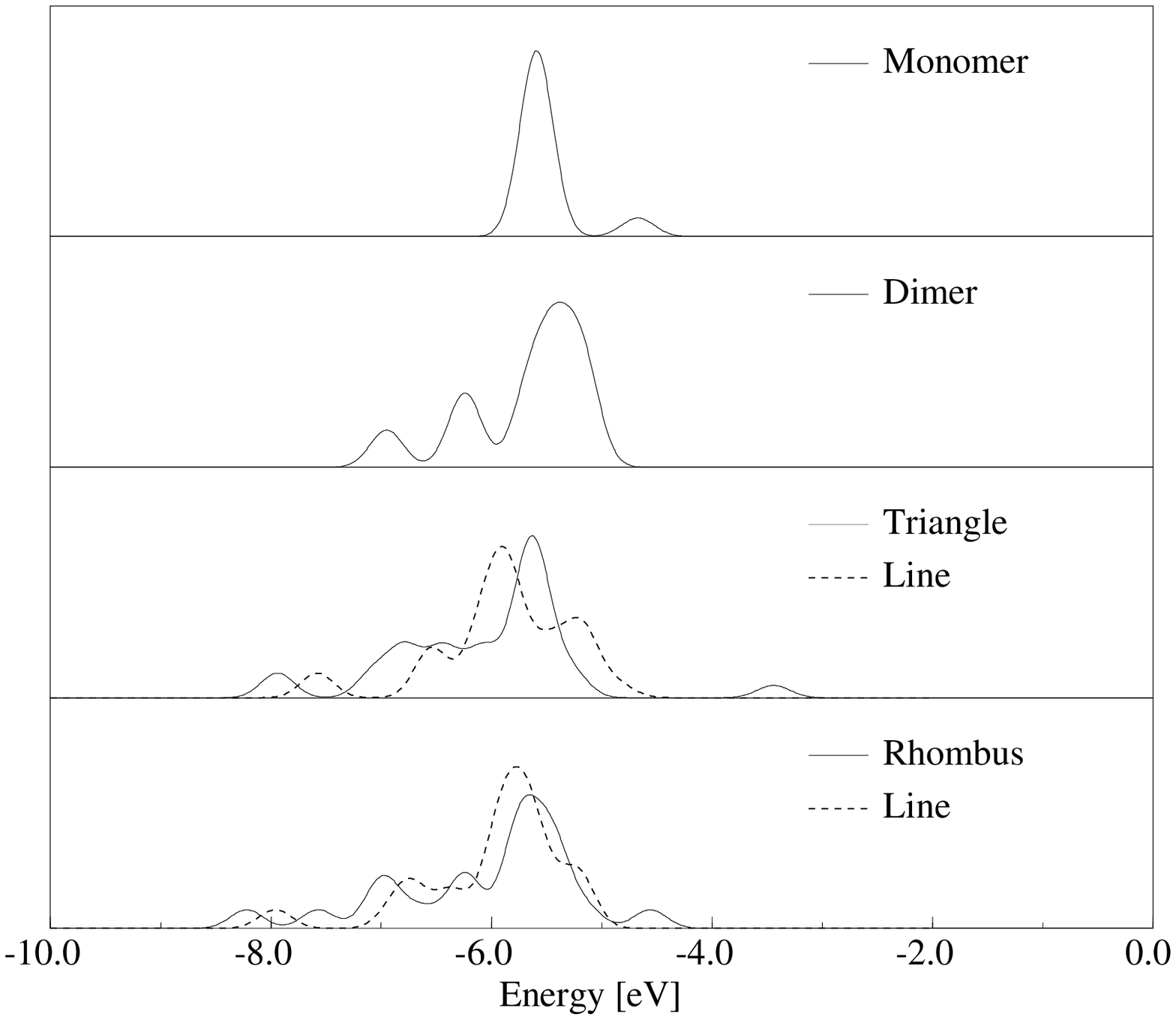,height=13cm}}
\vspace*{5cm}
\caption{Musolino {\em et al.}, Journal of Chemical Physics}
\label{fig:1-2-3-4}
\end{figure}

\capt{\protect\ref{fig:1-2-3-4}}{LDA density of states of free
clusters. All DOS have been scaled down so that total charge is unity.}

\begin{figure}[p]
\centerline{\psfig{figure=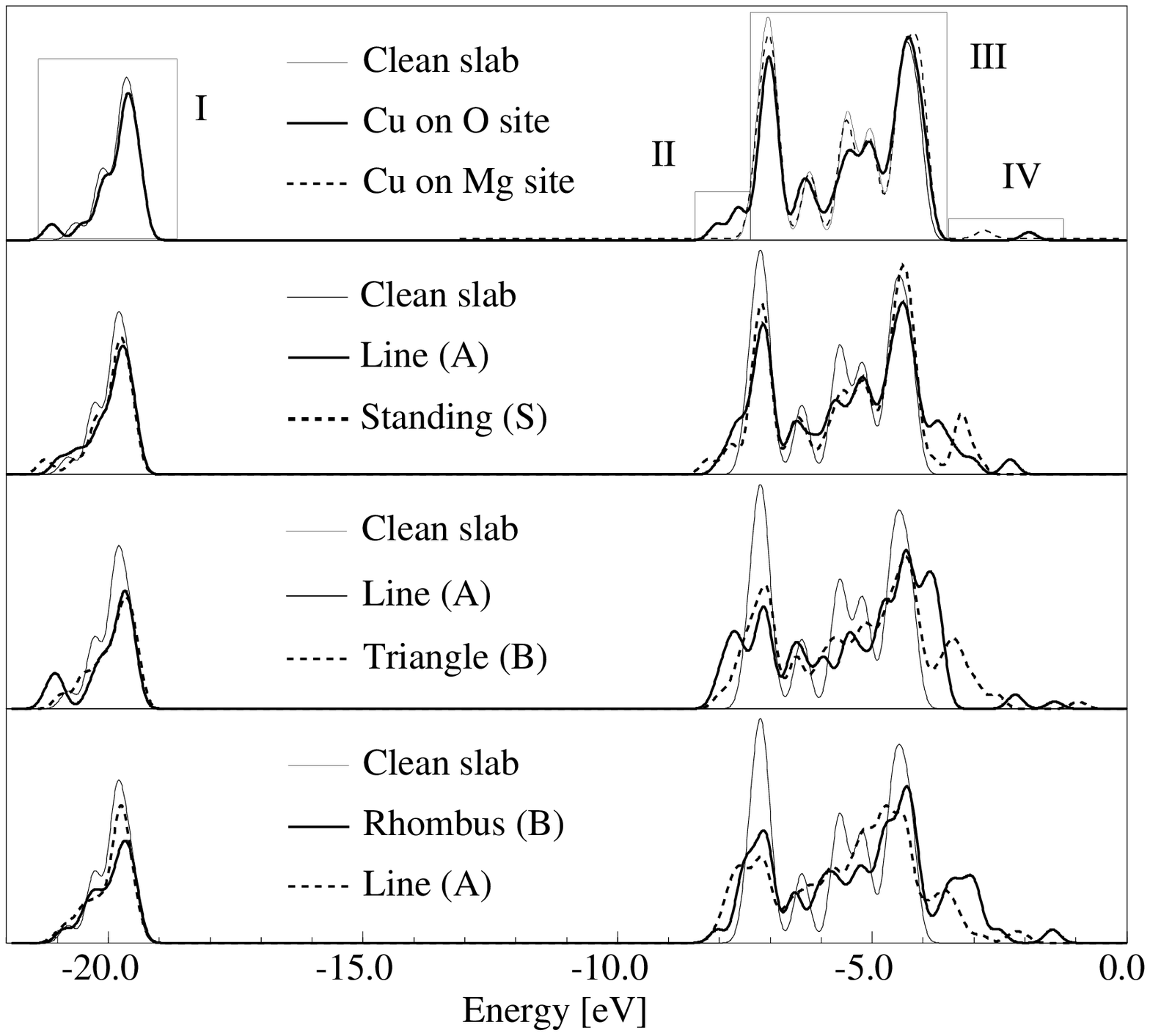,height=13cm}}
\vspace*{5cm}
\caption{Musolino {\em et al.}, Journal of Chemical Physics}
\label{fig:1-2-3-4_ads}
\end{figure}

\capt{\protect\ref{fig:1-2-3-4_ads}}{LDA density of states for the
adsorbed Cu$_{n}$ clusters ($n=1$ to $n=4$ from top to bottom). All
densities have been scaled down so that total charge is unity and
aligned on the top of the oxygen 2s band.}

\begin{figure}[p]
\centerline{\psfig{figure=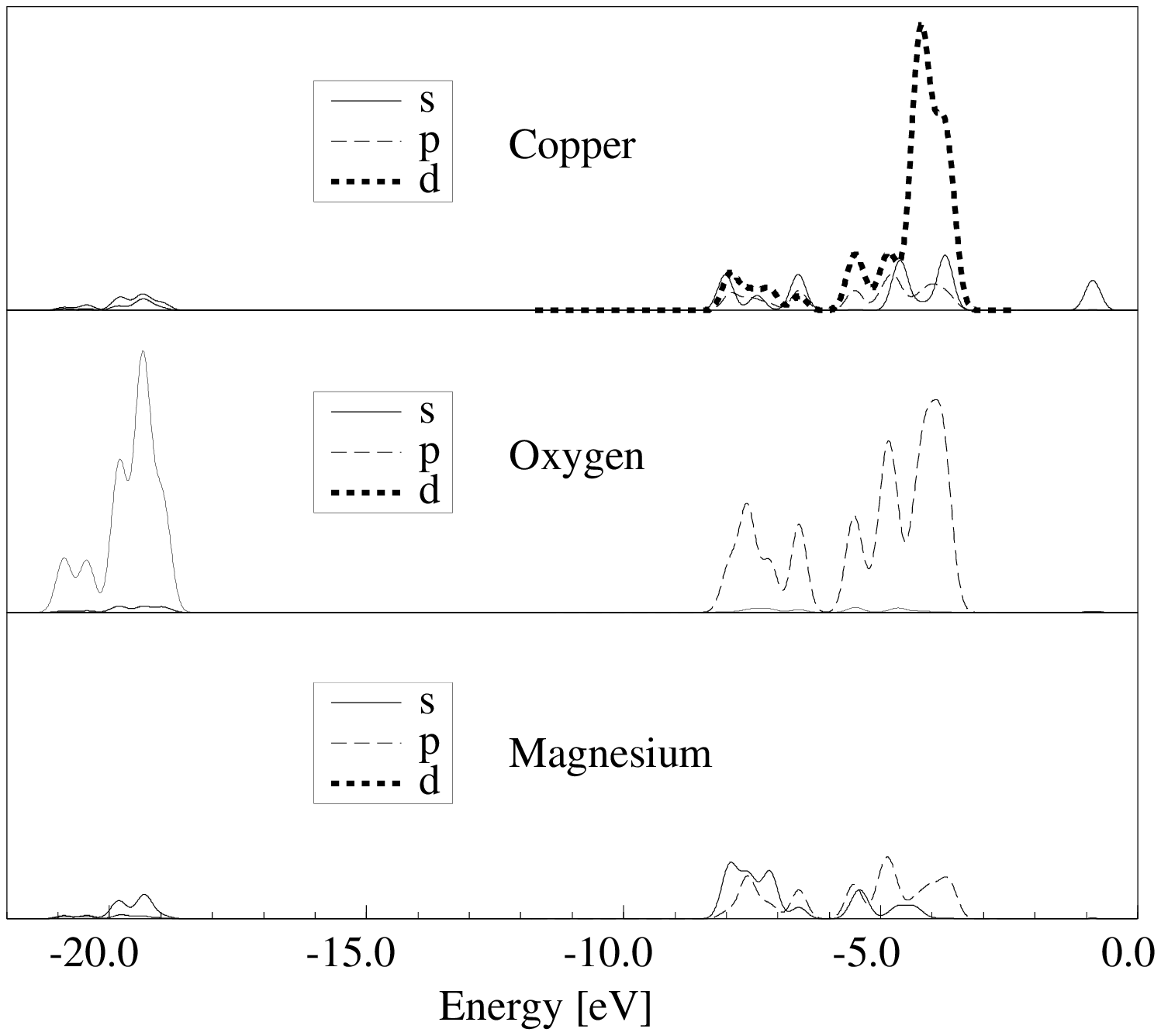,height=13cm}}
\vspace*{5cm}
\caption{Musolino {\em et al.}, Journal of Chemical Physics}
\label{fig:proj_dos}
\end{figure}

\capt{\protect\ref{fig:proj_dos}}{S, p, d projections of the DOS for a
Cu adatom adsorbed at the oxygen site.}

\begin{figure}[p]
\centerline{\psfig{figure=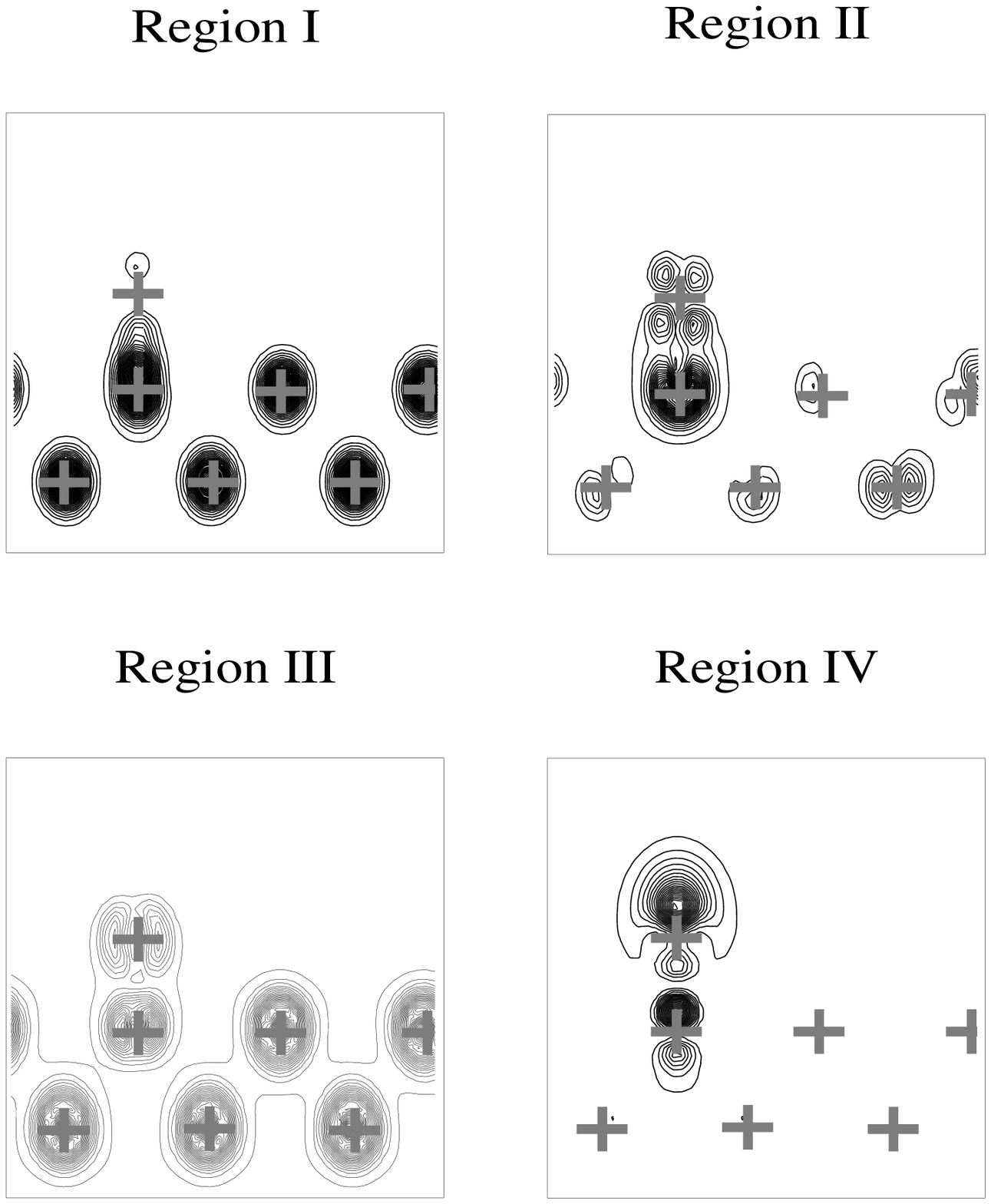,height=13cm}}
\vspace*{5cm}
\caption{Musolino {\em et al.}, Journal of Chemical Physics}
\label{fig:cd_all}
\end{figure}

\capt{\protect\ref{fig:cd_all}}{LDA charge densities for a Cu adatom
on top of an oxygen site. This view is along the (100) plane,
containing both oxygen and magnesium atoms of the slab. Crosses denote
oxygen and copper atomic positions.}

\begin{figure}[p]
\centerline{\psfig{figure=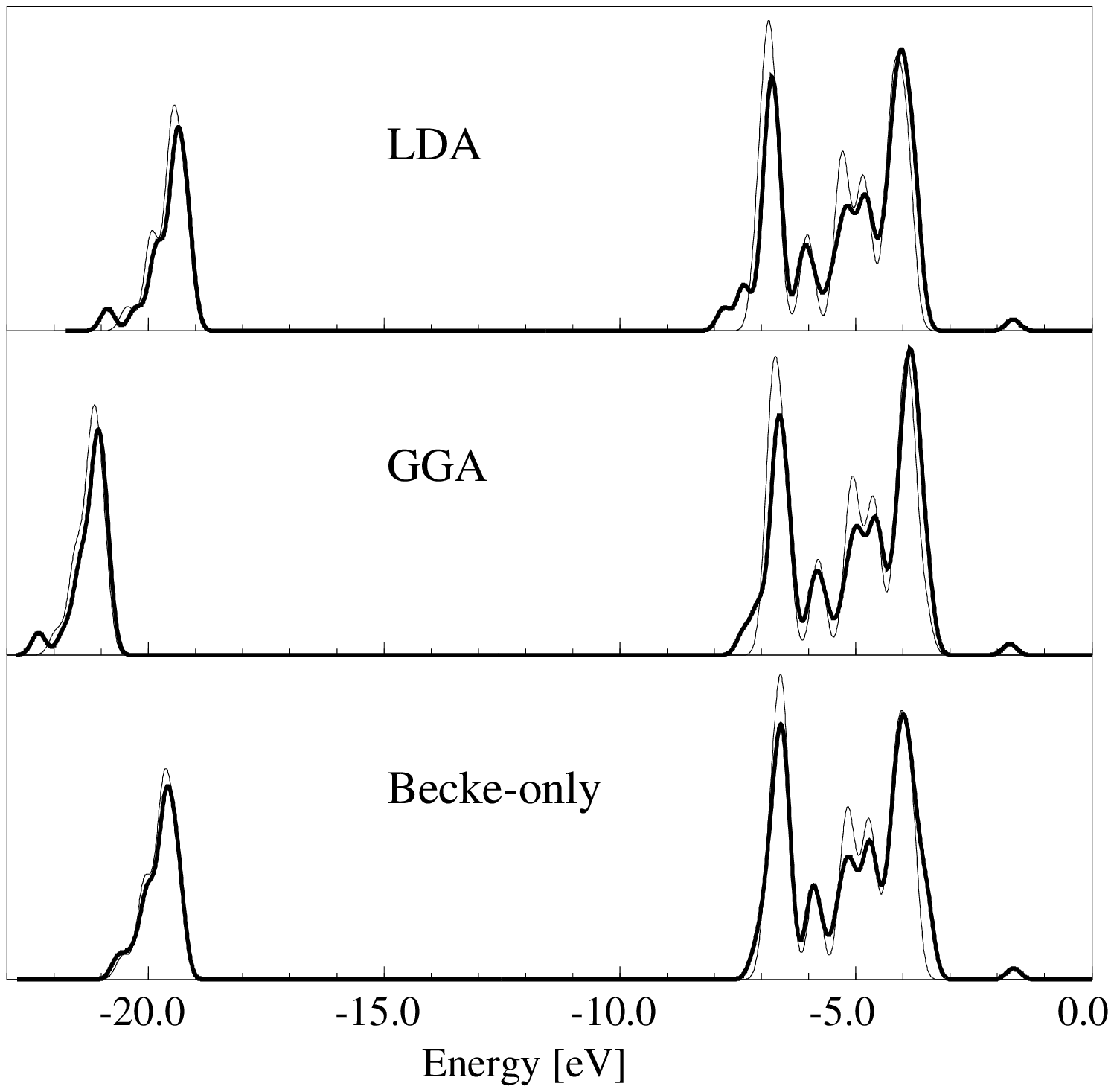,height=13cm,width=13cm}}
\vspace*{5cm}
\caption{Musolino {\em et al.}, Journal of Chemical Physics}
\label{fig:dos_comparison}
\end{figure}

\capt{\protect\ref{fig:dos_comparison}}{Density of states for a Cu
adatom at the oxygen site calculated with different
exchange-correlation functionals (bold line). All calculations are
fully selfconsistent and include the optimization of the atomic
structure. For comparison, the DOS of the clean surface (thin line) is
included. The DOS have been aligned on the top of the corresponding
clean surface's oxygen 2$s$ band.}

\newpage

\newpage

\newpage

\newpage

\newpage

\newpage

\newpage

\newpage

\newpage

\newpage

\end{document}